\documentstyle[prb,psfig,epsfig,aps]{revtex}
\input{epsf}
\begin{document}
\draft
\title{Incommensurate dynamics of resonant breathers in Josephson junction 
ladders}
\author{M. V. Fistul, A. E. Miroshnichenko and S. Flach}
\address{Max-Planck-Institut f\"ur Physik komplexer Systeme, N\"othnitzer
Strasse 38, D-01187 Dresden, Germany}
\author{M. Schuster and A. V. Ustinov}
\address{Physikalisches Institut III, Universit\"at Erlangen-N\"urnberg,
  D-91058 Erlangen, Germany}
\date{\today}
\wideabs{
\maketitle
\begin{abstract}
We present theoretical and experimental studies of resonant localized 
resistive states in a Josephson junction ladder. These complex breather states 
are obtained by tuning the breather frequency into the upper band of linear 
electromagnetic  
oscillations of the ladder. 
Their prominent feature is the appearance of
resonant steps in the 
current-voltage ($I$-$V$) characteristics. We have found the resonant 
breather-like states displaying {\it incommensurate} dynamics. 
Numerical simulations show that these incommensurate resonant breathers 
persist for very low values of damping.
Qualitatively similar incommensurate breather states are observed in
experiments performed with Nb-based Josephson ladders.
We explain the appearance of 
these states with the help of
resonance-induced hysteresis features in the $I$-$V$ dependence.
\end{abstract}
\pacs{74.50+r, 05.45Yv, 63.20Ls }
} 

A lot of attention has been recently devoted to the theoretical and experimental
study of {\it intrinsic dynamic localized states} in various
physical\cite{bisjabsplgfsapsarbwzwmis99,utslqeajs99}, chemical 
and biological systems \cite{sagk01,mp98}. These peculiar inhomogeneous 
states, called discrete 
breathers, appear in spatially homogeneous (Hamiltonian or dissipative) 
nonlinear discrete systems\cite{sa97,sfcrw98}. It was found that these 
states manifest themselves in 
the spectra of molecules and solids, and can determine the energy 
transfer in complex biological systems \cite{sagkammgpt01}.  
Discrete breathers have been observed e.g. in 
weakly coupled optical waveguides, in anti-ferromagnetic solids 
and in systems of 
interacting Josephson
junctions
\cite{etjjmtpo00,pbdaavusfyz00,pbdaavu00,lmfjlmpjmffsa96,sfms99,etjjmabtpo00}.
%
\begin{figure}[htb]
\vspace{10pt}
\centerline{\psfig{figure=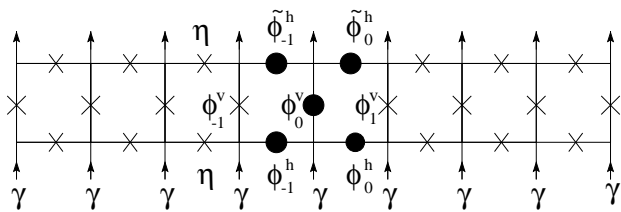,width=82mm}}
\vspace{20pt}
\caption{
Josephson junction ladder. Crosses mark the individual junctions. 
Arrows indicate the direction of external current flow (dc bias $\gamma$).
The black circles 
indicate the junctions that are in the whirling (resistive) state.
The breather state displaying up-down symmetry is shown,
}
\label{fig1}
\end{figure}
Especially in the latter example, namely {\it homogeneously} dc driven 
Josephson junction ladders (JJLs) 
(see Fig.~1),
an enormous diversity of breather states has been observed.
The JJL consists of small Josephson junctions of two types, 
"horizontal" and "vertical" ones, which are, respectively 
arranged perpendicular and parallel to the applied dc bias current $\gamma$. 
Breather states in a JJL are characterized by spatially localized 
voltage patterns as a few Josephson junctions are in the resistive (whirling) 
states (the black circles in Fig.~1) while the rest of the JJL junctions 
are in the superconducting state. 
The dynamics of breather states crucially depends
on two parameters: the anisotropy $\eta=\frac{I_{cH}}{I_{cV}}$, where 
$I_{cH}$ and $I_{cV}$ are respectively the critical currents of horizontal 
and vertical junctions, and the discreteness parameter 
(normalized inductance of the cell), $\beta_L$. 

The complete dynamics of JJLs is determined by the time dependent Josephson phases 
$\phi^v_i(t)$ (vertical), $\phi^h_i(t)$ (lower horizontal), and
$\tilde \phi^h_i(t)$ (upper horizontal),
and is governed by the Kirchhoff's laws and the flux quantization law. 
By making use of the resistively shunted model for Josephson junctions the 
following set of equations has been obtained \cite{gggfspug96}:
\begin{eqnarray}\label{2-8}
   \ddot{\phi}_{n}^v+\alpha\dot{\phi}^v_{n}+\sin\phi_{n}^v&
   =&\gamma+\frac{1}{\beta_{L}}(\triangle\phi_{n}^v+\nabla\phi_{n-1}^h
   -\nabla\tilde{\phi}_{n-1}^h)\nonumber \\
   \ddot{\phi}_{n}^h+\alpha\dot{\phi}_{n}^h+\sin\phi_{n}^h &
   =&-\frac{1}{\eta\beta_{L}}(\nabla\phi_{n}^v+\phi_{n}^h-\tilde{\phi}_{n}^h) 
   \\
  \ddot{\tilde{\phi}}{}^h_n+\alpha\dot{\tilde{\phi}}{}^h_n+
  \sin\tilde{\phi}_n^h&=&\frac{1}{\eta\beta_{L}}
  (\nabla\phi_{n}^v+\phi^h_n-\tilde{\phi}_{n}^h)\nonumber\;, 
\end{eqnarray}
where we use the notations $\triangle f_n\equiv f_{n-1}-2f_n+f_{n+1}$ 
and $\nabla f_n\equiv f_{n+1}-f_n$. Here, the unit of time is the inverse of 
the plasma frequency $\omega_p$, and $\alpha$ is the damping constant for 
a single junction. The dc voltage drop across a single junction is defined through a 
time average $<\dot \phi (t)>$.
Of crucial importance for the rich variety of breather states
is the presence of a nonzero dissipation in JJLs. This dissipation ensures that 
breathers can survive resonant interaction with extended 
linear electromagnetic waves (EWs), while their dynamic complexity
may actually increase at the same time.
Indeed, these complex {\it resonant} 
breather states have been predicted in
[Ref.~\onlinecite{etjjmtpo00}], experimentally observed in 
[Ref.~\onlinecite{mspbavu01}], and analyzed in detail in [Ref.~\onlinecite{aemsfmvfyzjbp01}]. 
The condition of the appearance of such a resonant breather 
state is the matching of the breather frequency $\Omega$ or 
its higher harmonics with the frequency 
$\omega(q)$ of one of the cavity modes of the ladder, $\omega(q)~=~m\Omega$. 
The breather frequency $\Omega$ 
is determined by the smallest nonzero dc voltage drop in the JJL, and the 
spectrum of cavity modes 
$\omega(q)$ has been derived previously using the linearization of 
the Eqs. (\ref{2-8}) around the superconducting ground state  
\cite{aemsfmvfyzjbp01,pcmvfavubamsf99}:

\begin{eqnarray}\label{2-13}
\omega_{\pm}^2&=&F\pm\sqrt{F^2-G}\;,\nonumber\\
F&=&\frac{1}{2}+\frac{1}{\beta_L\eta}+\frac{1}{2}
\sqrt{1-\gamma^2}+\frac{1}{\beta_L}(1-\cos q)\;,\\
G&=&(1+\frac{2}{\beta_L\eta})\sqrt{1-\gamma^2}+\frac{2}{\beta_L}(1-\cos
q)\nonumber\;.
\end{eqnarray}

For a finite size ladder with open boundary conditions, the spectrum of linear
waves is discrete and characterized by the following choice of
allowed wave number values:
\begin{equation}
q_l=\frac{l\pi}{N+1}\; ~ l=0,1,2...,N,
\label{obc}
\end{equation}
where $N$ is the number of cells.
Note here, that the resonant breather state is still a strictly {\it 
 periodic} solution 
in time. Its resonant character
is revealed by
large amplitude librations of the Josephson phases in the whole 
JJL.
Resonant breather states are manifested by resonant steps
in the current-voltage ($I$-$V$) characteristics, i.e. by a region where
the voltage saturates at a fixed value while the bias $\gamma$ is increased. 

The presence of 
the resonant breather state leads to the appearance of
{\it hysteresis} in  
the $I$-$V$ curve. Consequently there exist particular ranges of 
$\gamma$ where both resonant and non-resonant breathers can be stable.
However, these two types of breathers are characterized by different 
(incommensurate) frequencies $\Omega$, and moreover, the frequencies depend in 
a different manner on $\gamma$.
Thus, the question naturally appears: Is it possible for a breather
state to be both resonant and non-resonant at the same time in a JJL?
This is important because nondissipative systems support only 
time-periodic breathers, while dissipative systems typically need an 
external time-dependent drive to generate quasiperiodic dynamics. 
The idea here is that the spatial nonuniformity of the states 
(localized breathers) acts in a similar way to external time-dependent drives, 
so that even a dc bias is enough to obtain quasiperiodic (in time) states.

Here we will consider a region of parameters where the breather 
strongly interacts with the EWs.
We find that for low values of damping $\alpha$ 
a novel breather-like state with {\it incommensurate} (aperiodic) dynamics 
persists. This 
incommensurate resonant state can be viewed as the mixture of
resonant and non-resonant breathers in the JJL. Moreover, we also 
report on the 
experimental observation of this peculiar breather state 
in an extremely underdamped 
JJL. The incommensurate resonant breather-like state leads to
two successive voltage jumps in the $I$-$V$ curve.

To study the breather dynamics we perform direct numerical simulations of the 
set of equations (\ref{2-8}).
To establish a 
large dc current region where 
a strong resonant interaction of 
the breather state with the EWs occurs we choose the discreteness
parameter $\beta_L~=0.378$. 
The ladder consists of $N=10$ cells and the anisotropy parameter  
$\eta~=~0.49$. We impose open boundary conditions and use 4th order 
Runge-Kutta method for integration. 
The initial value of the dc bias was $\gamma=0.8$. 
We select proper initial conditions that lead to the relaxation of the system 
into a particular breather state of up-down symmetry 
with one resistive vertical junction, as shown in Fig.~1. Similar to our 
previous work\cite{aemsfmvfyzjbp01},
after a waiting time of 500 time units
we use the next 500 time units to calculate the time averaged characteristics
of the state. Then the dc bias $\gamma$ was decreased (increased) by a
step of $\Delta\gamma=0.001$ and  the 
procedure was repeated. We checked that our results do not change upon further
increase of the waiting time. 
The $I-V$
characteristics, that is the dependence of the averaged
voltage drop across the resistive Josephson junction on the dc bias, 
was monitored. The simulations were carried out for two different values of 
damping $\alpha~=~0.1$ (see Fig.~2) and $\alpha~=~0.025$ (see Fig.~3).

%
\begin{figure}[htb]
\vspace{20pt}
\centerline{\psfig{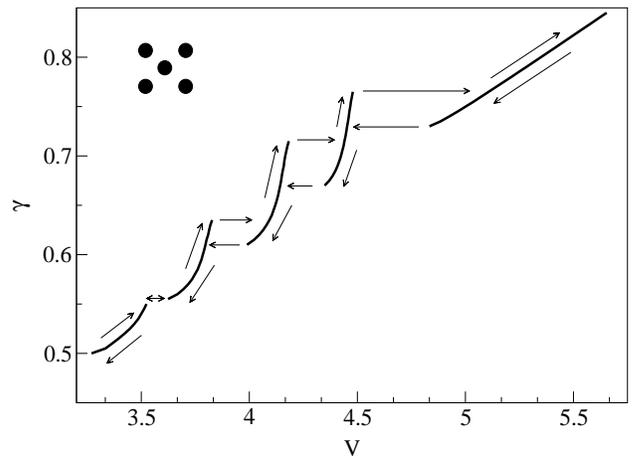}}
\vspace{2pt}
\caption{$I-V$ characteristics 
for $\alpha=0.1$, $\beta_L=0.378$, $\eta=0.49$. Arrows show the dc bias 
current increase (decrease). The dc voltage pattern of the breather state 
is presented by the black circles.}
\vspace{10pt}
\label{fig2}
\end{figure}

In the case of a moderate value of damping $\alpha~=~0.1$ 
we obtain a non-resonant breather state for dc bias current values down to
$\gamma~\geq~0.72$. This breather state displays a 
complete up-down symmetry, $\phi^h_i(t)~=~-\tilde \phi^h_i(t)$. 
The $I$-$V$ curve is linear in this regime. 
In the region of moderate dc current bias values 
resonant steps appear in the $I$-$V$ curve (Fig.~2). The presence of 
these steps indicates a strong resonant interaction between the 
breather state and EWs as the condition $\omega_+(q_l)~=~2\Omega$ is satisfied. 
The observed steps correspond to resonances with cavity modes with 
particular values of $l~=~2, 4, 6, 8$.
As the JJL is biased at the steps, a resonant type of the 
breather state is realized.  This state conserves the structure of the 
dc voltage pattern but is characterized by a large ac voltage component 
of the superconducting Josephson junctions.  
Independent of the observed dynamical complexity, 
all breather states (resonant or 
non-resonant type) found for this value of damping $\alpha$ display a 
time-periodic behavior. 

However, the situation changes drastically as we lower the  
dissipation down to $\alpha~=~0.025$.  
First, the breather state strongly interacts with EWs if the condition 
$\omega(q_l)~=~\Omega$ is matched. Secondly, we observe only a single resonant 
step corresponding to the particular cavity mode ( $l=2$) excitation. 
Under these conditions we found that the switching from 
the non-resonant breather state\cite{comment} 
appearing for large values of dc voltages,
to the 
resonant breather occurs in the form of two successive voltage jumps 
(see, Fig.~3a). 
%
\begin{figure}[htb]
\vspace{20pt}
\centerline{\psfig{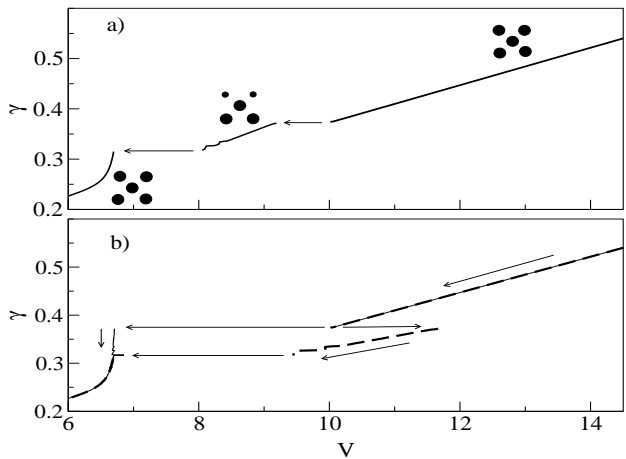}}
\vspace{2pt}
\caption{$I-V$ characteristics 
for $\alpha=0.025$, $\beta_L=0.378$, $\eta=0.49$. Arrows indicate the 
voltage jumps and show the decrease of the dc bias current. 
a) Result for voltage drop 
across the vertical junction. The size of the black circles indicates the relative values of 
dc voltage drops. b) Result for twice the voltage drop across the upper 
(solid line) and lower (dashed line) horizontal junctions. 
}
\vspace{10pt}
\label{fig3}
\end{figure}
Thus, in the particular region of the dc bias 
current 
($0.31~<~\gamma~<~0.38$) the up-down symmetry of the dc voltage pattern 
is violated 
and a novel resonant breather-like state with incommensurate (aperiodic) 
dynamics appears. Indeed, as the system jumps to this state 
there are at least two incommensurate frequencies, 
namely $\Omega_1~\simeq~3.4$ and $\Omega_2~\simeq~5.3 $.
Moreover, with decreasing dc bias the value of $\Omega_1$,
practically does not change, in contrast to $\Omega_2$.
A closer look at the $I$-$V$ curves of the 
horizontal junctions (Fig.~3b) also shows that the upper (lower) horizontal 
junctions display the resonant (non-resonant) behavior. Thus
we can conclude that this novel incommensurate 
breather-like state can be viewed as a coexistence of resonant and 
non-resonant junction behavior in the JJL. We note here that similar aperiodic 
breather states were found in a wide range of parameters $\beta_L~\simeq~1$ and 
$\eta~\leq~1$. 

By making use of a dc analysis~\cite{etjjmtpo00,aemsfmvfyzjbp01}, i.e. 
neglecting ac components of the Josephson current, 
we find that the
voltage jumps $\Delta V_h$ and $\Delta \tilde V_h$
of the non-resonant and resonant horizontal junctions have to satisfy
the relationship:
\begin{eqnarray}
\Delta V_h~=~\frac{\Delta \tilde V_h}{1+2\eta}
\end{eqnarray}
Moreover, the presence of the incommensurate resonant breather state is 
characterized by a lower resistance $\propto(\alpha (2\eta+1))^{-1} $ 
as compared to the non-resonant case.
Indeed, our simulations show that the dc voltage jump of the lower horizontal 
junctions is approximately two times smaller than the corresponding dc voltage 
jump of the 
upper horizontal junctions, in full accordance with the dc analysis.

In order to characterize an incommensurate breather state more precisely 
we calculate the Fourier transform of the time dependent 
voltage $\dot \phi_1^v(t)$ of the vertical junction (see Fig. 4), 
which is in the superconducting state and the closest one 
to the core of the breather. 
\begin{figure}[htb]
\vspace{20pt}
\centerline{\psfig{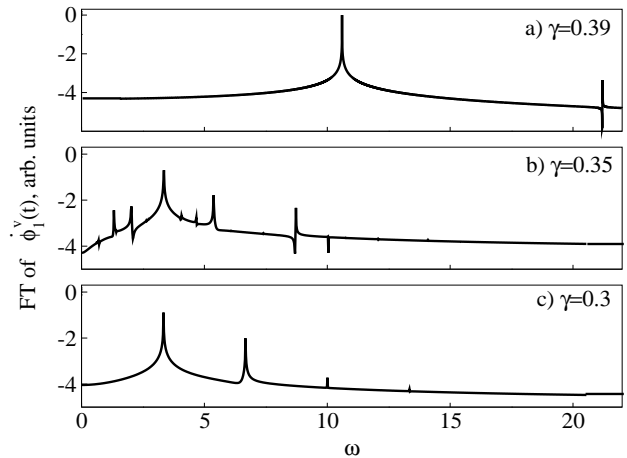}}
\vspace{2pt}
\caption{The Fourier transform (FT) of the ac voltage $\dot \phi_1^v(t)$. 
Three different values of dc bias current are shown and 
the logarithmic scale for FT was used.}
\vspace{10pt}
\label{fig4}
\end{figure}
If the breather state is the periodic one , i. e. 
for large (Fig. 4a) and small (Fig. 4c) values of dc bias current, 
the Fourier transform contains the breather frequency $\Omega$ and higher
harmonics. However, in the intermediate region of the dc bias 
as the breather displays incommensurate dynamics, the Fourier 
transform becomes more rich (Fig. 4b). We observe two different harmonics 
$\Omega_1$ and 
$\Omega_2$, and the various combination frequencies $m\Omega_1 ~+~n\Omega_2$, 
where $n$ and $m$ are integers.

The incommensurate breather-like states analyzed above were
initially observed in experiments which we performed using
Nb-based Josephson junction ladders.
The experimental layout is similar to the
one presented in [Ref. \onlinecite{pbdaavusfyz00}] and corresponds to the 
schematic shown in Fig.~1. The measurements were
performed in liquid helium at a temperature of
$T=4.2\,\mathrm{K}$. The data we present were taken from a 10 cell
open-ended Josephson ladder of anisotropy $\eta=0.49$. 
By
chosing a relatively low critical current density of
$j_c\approx100\;\mathrm{A/cm^2}$ for the samples we achieved a discreteness
parameter of $\beta_L=0.62$, which makes it possible for a breather to be
strongly resonant with EWs. At the same time, the low damping value
of $\alpha=0.025$ was sustained.

The breather state was created artificially by 
initially applying a local bias current $\gamma_l$ to one
of the vertical junctions, and thus, forcing the switching of this junction 
into the resistive state. After that we decreased $\gamma_l$ and simultaneously 
increased the homogeneous bias current $\gamma$. This procedure, described 
in detail in [Ref. \onlinecite{pbdaavusfyz00}], finally yields the desired breather
state in the presence of an uniformly supplied bias current.
In this paper we focused on an up-down symmetric breather state with dc 
voltage pattern shown in Fig.~1.  
The $I$-$V$ curves were obtained by sweeping the bias current $\gamma$  
and measuring the respective voltage drops across the junctions. For
convenience, we plot normalized voltage $v=V/V_0$ with
$V_0=\frac{\Phi_0}{2\pi}\omega_p=84.3\,\mu V$, and normalized bias current 
$\gamma=I_{ext}/I_{cV}$ with $I_c=20.5\,\mu A$. 

For the sample \#1
the breather state was created at a dc bias current $\gamma~=~0.38$.
The acquired $I$-$V$ curve, as shown in Fig.~5, displays a
linear, non-resonant behavior until $\gamma~= ~0.35$ (see Fig.~5).
The breather state 
is symmetric, and the absolute values of the horizontal junction voltages
$|V_h|$ and $|\tilde V_h|$
are just one half of the vertical junction voltage, $|V_v|$. 
As the dc bias is 
decreased further, the switching to the resonant breather state takes place. 
This switching occurs through an intermediate state (see
Fig.~5a). Indeed, in this region of $0.32<\gamma<0.35$ we  
observe the symmetry-broken state with $|\tilde V_h|\ne|V_h|$. 
Moreover, the ratio of voltages $|\tilde V_h|/|V_h|$ continuously changes 
as $\gamma$ decreases (see Fig.~5b). For the values of $\gamma~<~0.32$ the 
symmetric breather state recovers but is now a resonant breather state.

For the sample \#2 we  observe a similar intermediate state 
appearing in the process of the switching from the non-resonant to the resonant 
breather state (see Fig.~6). 
In this experiment carried out 
in the presence of a small magnetic field 
(the normalized magnetic flux per cell 
was less than 0.1), the $I$-$V$ 
curve of the non-resonant (lower) horizontal junction (see Fig.~6b) displays a 
voltage jump similar to the one found in the numerical studies (see Fig.~3). 
The ratio of the voltages and the corresponding rotation 
frequencies of the horizontal (upper and lower) junctions increases as 
$\gamma$ decreases. 
The measured $I$-$V$ curves in both cases qualitatively look similar to 
the current-voltage characteristics obtained by numerical simulations of (1).
Thus, the predicted breather state displaying  aperiodic dynamics  is clearly 
observed (compare Figs.~3 and 5,6).
%
\begin{figure}
  \begin{center}
    \epsfig{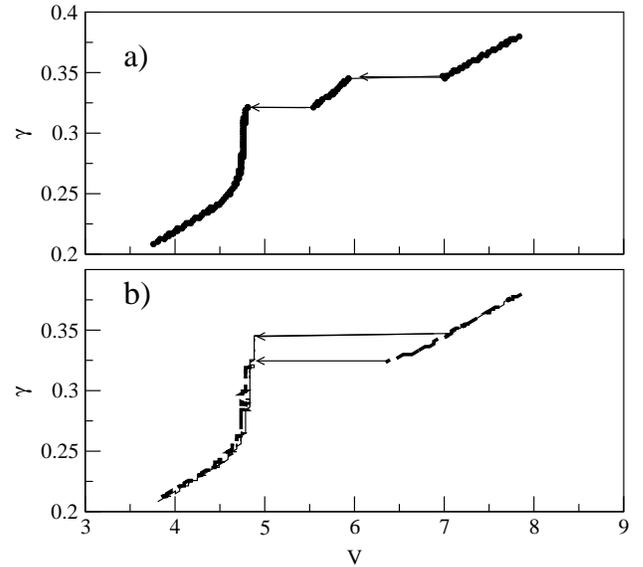}\\[.3cm]
    \caption{Measured $I$-$V$ curves (sample \#1) of an initially symmetric
      breather state created at a dc bias of $\gamma=0.38$: (a) 
      voltage drop across vertical junction,
      (b) voltage drops across upper (thin solid line) 
      and lower (dashed line) left horizontal junctions. The breather state 
      with incommensurate dynamics  is formed for
      $0.32<\gamma<0.35$. The fine structure superimposed on the curve
      is due to reduced experimental resolution in these measurements.
       }
    \label{fig5}
  \end{center}
\end{figure}
However,  there are several differences between the experimental observations and 
the numerical analysis of (1). Firstly, in contrast to numerics 
we did not observe the voltage jump of the non-resonant horizontal junction in 
the measurement shown in Fig.~5, and secondly, in Fig.~6 we observe
just a
tiny voltage difference between upper and lower horizontal junctions
in the bias region above the incommensurate state. This particular
behavior is not yet explained.
We want to stress that the incommensurate breather state was observed
under various experimental conditions, which were partly influenced
by magnetic fields in the vicinity of the studied
sample.

In conclusion, we found an inhomogeneous resistive state showing
aperiodic dynamical behavior by making use of a numerical analysis of
(1).
Similar incommensurate breather states were observed experimentally 
in niobium-based underdamped Josephson junction ladders.
The incommensurate breather states are characterized by a resonant 
behavior of some horizontal junctions and a non-resonant behavior of
the other horizontal junctions. Such states appear in the process 
of a switching from symmetric non-resonant to symmetric resonant breather state.
The origin of these particular breather states is a  
strong resonant excitation of EWs 
by the breather state leading to the resonance-induced hysteresis in the 
$I$-$V$ curve.  Thus, we argue that the aperiodic
inhomogeneous dynamic states can present a generic feature of 
dissipative (but extremely underdamped) lattices.
%
\begin{figure}
  \begin{center}
    \epsfig{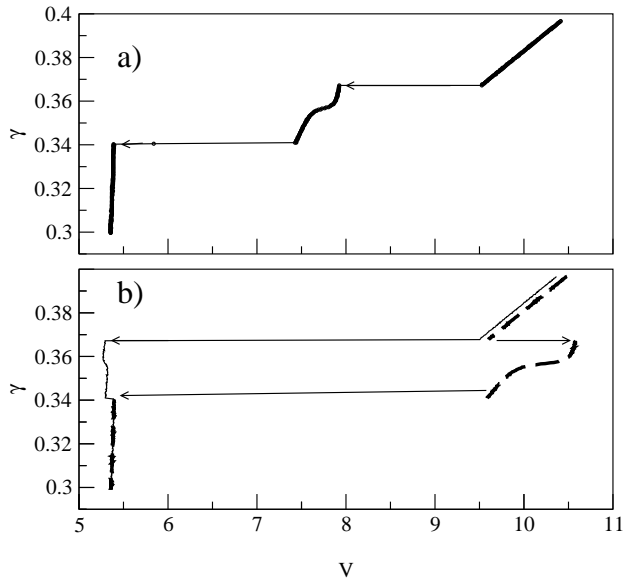}\\[.3cm]
    \caption{Measured $I$-$V$ curves (sample \#2) of an initially symmetric
      breather state created at a dc bias of $\gamma=0.4$: (a) 
      voltage drop across vertical junction,
      (b) voltage drops across upper (thin solid line) 
      and lower (dashed line) left horizontal junctions. The breather state 
      with incommensurate voltages is formed for $0.34<\gamma<0.37$. 
      }
    \label{fig6}
  \end{center}
\end{figure}

We thank A. Benabdallah for useful discussion. This work was supported by the Deutsche Forschungsgemeinschaft
and
by the European Union under the RTN project LOCNET HPRN-CT-1999-00163.


\begin{references}
\bibitem{comment}In the case of a small dissipation we also found that 
a non-resonant breather state displays a broken dynamic symmetry. Thus,
the breather state with a complete up-down symmetry becomes unstable and 
the stable (non-resonant) states are characterized by a relationship:
$\langle\phi^h_0 (t)+\tilde{\phi}^h_0 (t)\rangle=\pm\pi$. The properties of these
peculiar breather states will be reported elsewhere.
\end{references}
\end{document}